\documentclass[doublecol]{epl2}

\usepackage{graphicx}
\usepackage{latexsym} 
\usepackage[centertags]{amsmath} 
\usepackage{amsfonts} 
\usepackage{amssymb} 
\usepackage{amsthm} 
\usepackage{newlfont} 
\usepackage{subfigure} 
\usepackage{amsbsy} 
\usepackage{color}
\usepackage{bm}

\def\k{{\mathbf k}}
\def\r{{\mathbf r}}
\def\v{{\mathbf v}}
\def\r{{\mathbf r}}

\def\hk{\hat{k}}

\title{Non-equilibrium length in granular fluids:\\ from experiment to fluctuating hydrodynamics}

\author{G. Gradenigo, A. Sarracino, D. Villamaina and A. Puglisi}
\institute{CNR-ISC and Dipartimento di Fisica, Universit\`a Sapienza - p.le A. Moro 2, 00185, Roma, Italy}

\abstract{Velocity correlations in a quasi-2D driven granular fluid are studied in
experiments and numerical simulations.  
The transverse velocity structure factor reveals 
two well defined energy scales,
associated with the external ``bath temperature'' $T_b$ and with the
internal granular one, $T_g<T_b$, relevant at large and small
wavelengths respectively.  Experimental and numerical data are
discussed within a fluctuating hydrodynamics model, which allows one
to define and measure a non-equilibrium coherence length,
growing with density, that characterizes order in the velocity field.}


\begin{document}

\maketitle

The study of collective phenomena in complex systems is a central
issue in statistical mechanics.  In particular, the degree of
\emph{order} emerging in such systems is customarily investigated by
studying the behaviour of correlation functions~\cite{F75}.  However,
in many systems which display interesting collective phenomena
interactions are \emph{non-conservative} and a stationary state is
achieved through a continuous and homogeneous energy injection
 acting on every single particle: this is the case, for instance, of
homogeneously vibrated granular materials as well as of the so-called ``active
fluids'' including swarms of bacteria, birds, insects, fishes,
pedestrians and even artificial motors~\cite{JKPJ07,NRM07,Cetal10}. In
all those systems, the study of correlations requires tools beyond the
standard equilibrium ones.

The emergence of order out of equilibrium is studied here in a
granular system~\cite{JN92,DMB05}, where inelastic collisions induce
cross correlations between particles velocities.  The behaviour of
such correlations is customarily described with fluctuating
hydrodynamics (FHD), which relies on a separation of time and length
scales in the system, together with a suitable model for
noise. Assessing the limits of validity of granular hydrodynamics is
non-trivial~\cite{K99,G99}, but in several situations it has proven to
be useful in explaining the behaviour of large scale correlations in
granular fluids~\cite{GZ93,NEBO97}. {In a more general context, the
existence of non-trivial correlations emerging from non-equilibrium
assumptions was already noticed in~\cite{GLS90}.} Recently, FHD also provided a
prediction for the slowing-down of the dynamics, with increasing
packing fraction, in granular fluids~\cite{KSZ10,VAZ11}, a phenomenon
which has a well known counterpart in elastic systems~\cite{BGL89}.
The above mentioned and similar results~\cite{SVGP10}, which clearly 
show the existence of a time-scale growing with the density, suggest to look also 
for a growing length-scale.

In this paper we show the existence of a finite correlation length
$\xi$ in the velocity field, peculiar to a non-conservative fluid.
Velocities correlations are investigated studying the velocity
structure factor measured from experiments and numerical
simulations. Data are discussed within a FHD model, allowing us
to extract a length scale $\xi$ and confirming that its associated
order is a peculiarity of the out-of-equilibrium regime.  The
behaviour of $\xi$ is investigated at different packing fractions,
showing a remarkable increase of ordering in the velocity field with
growing density. The large scale fluctuations of the
velocity field suggest that a reference
theoretical model where the thermostat has a finite viscous
drag~\cite{PLMPV98} is the most appropriate to modelize the
experiments. 

\begin{figure}[!htb]
\includegraphics[width=1.0\columnwidth,clip=true]{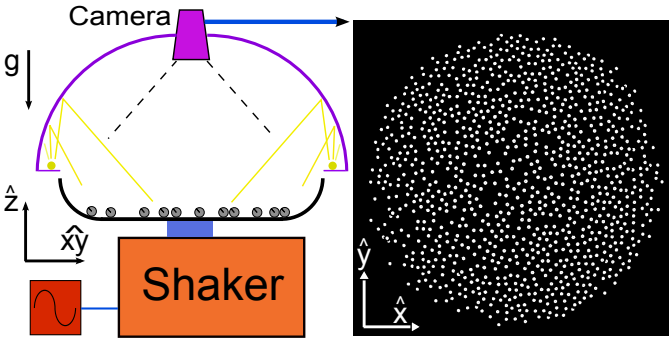}
\caption{(color). Schematic of the experimental setup.}
\label{setup}
\end{figure}

The experimental setup, sketched in Fig.~\ref{setup}, consists of an
electrodynamic shaker (LDS V450) which vibrates vertically ($\hat{z}$
axis) a horizontal rigid aluminium plate covered by a partial granular
monolayer of $N$ ``steel 316'' spheres, with packing fraction
$\phi=\pi n\sigma^2/4<0.5$, where $n=N/Area$ is the number density and
the diameter $\sigma$ of spheres is $4$ mm.  The plate is circular and
has slowly rising boundaries in order to prevent highly dissipative
head-on collisions between spheres and walls. The radius of the
central flat region is $R=100$ mm.  The surface of the plate is
rough~\cite{PEU02,RIS07a,RIS07b}, transferring part of the vertical
momentum to the horizontal ($\widehat{xy}$ plane) motion of the
spheres. {The plate is roughened by stainless-steel
  sand-blasting with a granulometry in the range $[100-500]\mu m$.}
Collisions among the spheres are also responsible for $\hat{z} \to
\widehat{xy}$ energy conversion. From a fast camera (Mikrotron EoSens
CL, up to $506$fps for $1280x1024$ pixels), placed on the top of the
plate, we observe the $\widehat{xy}$ motion of the
spheres. {Even if the container has no lid (as it was in the
  experiment with the rough plate in~\cite{PEU02}), we verified by
  visual inspection that - at the chosen shaking parameters (see
  below) - particles never overcome each other and constitute a
  quasi-two-dimensional fluid.}
The results discussed below are obtained for sinusoidal shaking,
i.e. the plate follows the law $z(t)=A\sin(\omega t)$, with
$A\omega^2/g=12$ and $\omega/(2\pi)=150$ Hz, where $g$ is the gravity
acceleration, yielding a maximum velocity of the vessel $v_m\simeq
125$ mm/s. A particle tracking software identifies particles' centers
of mass, up to an accuracy of $1/5$ of a pixel, and their displacement along close successive frames ($\Delta
t=1/57$ sec), reconstructing particles' velocities: the analysis of
the mean squared displacement $\Delta s^2(\tau)=\langle |x(t+\tau)-x(t)|^2
\rangle$ guarantees that the reconstruction is operated in the
ballistic regime, i.e. $\Delta s^2(\tau) \sim \tau^2$ for $\tau < 2 \Delta
t$.

\begin{figure}[!tb]
\includegraphics[width=1.0\columnwidth,clip=true]{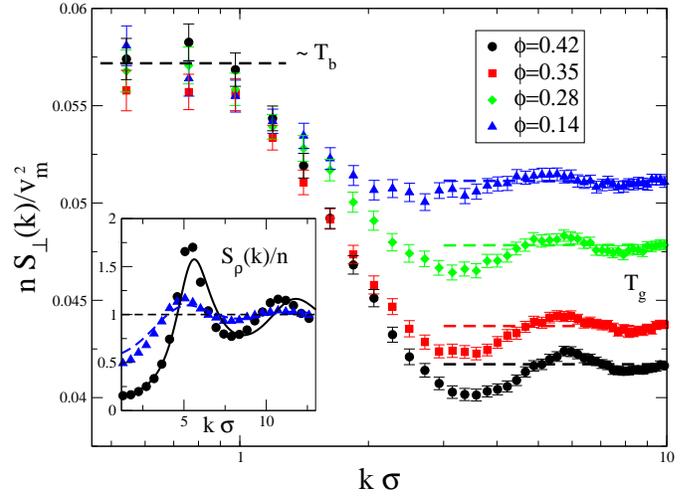}
\caption{(color online). $S_\perp(k)$ measured in units of $v_m^2$ in
  experiments {(where $v_m$ is the maximum velocity of the
    vibrating plate)}, for different values of $\phi$.  Data represent
  averages over 3300 pairs of frames.  The dotted lines on the right
  show the corresponding $T_g$. {As a reference for the eye,
    we also indicate the average (among the four experiments) of
    $T_b$, which is defined in the text.}  Inset: $S_\rho(k)$ for
  $\phi=0.14$ ({\color{blue}$\blacktriangle$}) and $\phi=0.42$
  ($\bullet$) with the analytical predictions for elastic
  disks~\cite{BC86}. In all figures the lowest wave-vector is $k\sigma
  \geq 2 k_{min}\sigma$ ($2 k_{min}\sigma = 4\pi\sigma/R=0.5$),
  because at smaller ones finite size-effect are dominant.  The values
  of $S_\perp(k)$ reported are an average on moments with modulus $k
  \in [n 2\pi/R , (n+1) 2\pi/R]$ for integer $n \geq 2$.}
\label{Skexp}
\end{figure}

The measure of the static structure factor $S_\rho(k) = 1/V
\sum_{i=1}^{N} \left\langle | \exp(-i \k \cdot \r_i) |^2
\right\rangle$, where $\r_i$ denotes the position of each particles
and $V$ is the volume, reveals that in the range of packing fraction
that we studied, i.e. $\phi \in \left[ 0.1,0.42 \right]$, the
structure of the granular fluid presents small differences from its
elastic counterpart (inset of Fig.~\ref{Skexp}), probably due to weak
clustering or inhomogeneities. {The small enhancement of the
  structure may also be understood by realizing that - in our setup - a
  slight amount of vertical motion allows particle to arrange at
  separations even smaller than their diameter. }

More peculiar is the behaviour of the velocity structure factor.
In particular we study here its transverse part:
\begin{eqnarray} 
n^2 S_\perp(k) 
&=&V^{-1} \langle v_\perp(k) v_\perp(-k)\rangle,
\label{eq:decomp}
\end{eqnarray}
with $v_\perp(k)=\sum_i\left(\v_i\cdot \hk_\perp\right)\,
\exp(-i\, \k\!\cdot\!\r_i)$, where $\v_i$ denotes the velocity of
the $i$-th particle and $\hk_\perp$ is the unitary vector such that
$\hk_\perp\cdot \hk=0$.
The curves of $nS_\perp(k)$ measured in the range of 
packing fractions $\phi \in [0.1;0.42]$, i.e. in a moderate dense fluid regime,
are shown in the main frame of Fig.~\ref{Skexp}.

By definition, $S_\perp(k)$ yields a measure of the energy localized
on average on each mode $k$.  In the granular system studied here, we
find that a different amount of energy is concentrated on each mode,
namely at stationarity $S_\perp(k)$ is a non trivial function of $k$.
More specifically, we observe that \emph{two} relevant energy scales
are well defined in the system: one at large $k$, which corresponds to
the granular temperature $T_g=1/(2N)\sum_i\langle v_i^2\rangle$ and
strongly depends on the packing fraction $\phi$; the other, at small
$k$, is weakly dependent on $\phi$.  Consider that in the limit $k\to
0$, we have that, by definition, $n S_\perp(k) \rightarrow N \langle
V_{cm}^2 \rangle$, with $V_{cm}$ the velocity of the center of mass of
the system.  The quantity $N \langle V_{cm}^2 \rangle$ is not affected
by momentum-conserving collisions: it is, therefore, related
\emph{only} to the interaction with the vibrating plate.
{For such a reason, we use this value as a measurement of the
  so-called ``bath temperature'' $T_b$.} In the theoretical model
introduced below this energy corresponds to the temperature of the
thermostat coupled to the granular fluid. From Fig.~\ref{Skexp} it is
evident that the gap between the two energy scales defined above
significantly grows as the packing fraction is increased. We will show
that the existence of such a gap is deeply connected with the
existence of a finite correlation length in the velocity
field. {The oscillations of $S_\perp(k)$ at high values of
  $k$, are likely to be a signature of an unexpected coupling with the density
  modes, already observed in previous works~\cite{NETP99}}.

With the aim of giving a consistent interpretation of our experimental
results, we now propose an effective model for the microscopic
dynamics of particles (restricted to the plane $xy$) and deduce
from it a theoretical formula - through fluctuating
hydrodynamics~\cite{NETP99,BMG09} - for $S_\perp(k)$.
Our proposal is well known in the literature~\cite{PLMPV98}: it mimics the action of the
rough vibrating plate by means of an interaction with an effective viscous ``bath''.
The model is described by the following equation governing the dynamics of the $i$-th particle ($i\in[1,N]$)
\begin{equation}
\dot{{\mathbf v}}_i(t)=-\gamma_b {\mathbf v}_i(t) + \boldsymbol{\zeta}_{i}(t)+{\mathbf F}_i,
\label{langgas}
\end{equation}
where ${\mathbf F}_i$ represents the effect of particle-particle
inelastic hard core collisions with restitution coefficient $\alpha$,
while $\gamma_b^{-1}$ is the typical interaction time with the plate
and $\boldsymbol{\zeta}_{i}(t)$ a white noise, with zero mean and
$\langle\zeta_{i,\alpha}(t)\zeta_{j,\beta}(t')\rangle
=2T_b\gamma_b\delta_{\alpha\beta}\delta(t-t')\delta_{ij}$
($\alpha,\beta=\{x,y\}$).  This kind of thermostat is well defined
also in the elastic limit $\alpha \to 1$, where the fluid equilibrates to the bath
temperature: $T_g=T_b$~\cite{PLMPV98}.

For the above model one can apply the usual kinetic
theory to study hydrodynamic fields~\cite{NETP99}.
In linearized hydrodynamics, which applies 
to the moderately dense regime here studied,
the transverse modes $v_\perp(k,t)$ of the velocity field
are decoupled from all the other modes.
In particular, at small $k$ they are well described - on average - 
by the equation $\dot{v}_\perp(k,t) = - (\gamma_b + \nu k^2 ) v_\perp(k,t) $,
with $\nu$ the kinematic viscosity~\cite{F75}. 
This equation contains two main sources of relaxation: the external friction
parametrized by $\gamma_b$, and vorticity
diffusion parametrized by $\nu$. To analyze structure factors one
needs to plug in appropriate noises.  The two relaxation mechanisms
described above are associated with two independent sources of
randomness: external and internal noise, respectively. A rigorous
evaluation of the way those noises enter the description for
$v_\perp(k,t)$ can be found in the literature for different
models~\cite{BMG09}. Here we present a more phenomenological approach,
assessed by its ability to reproduce experimental and numerical
results. The idea, partly inspired to previous granular FHD
theories~\cite{NETP99}, consists in keeping valid - as a first
approximation - the $2$nd kind Fluctuation Dissipation Relation (FDR)
for the two relaxation-noise pairs, separately. The result can be cast
in a Langevin equation for the shear mode, with one single noise with
appropriate variance:
\begin{eqnarray}
\dot{v}_\perp(k,t) &=& - (\gamma_b + \nu k^2 ) v_\perp(k,t) 
+ \eta(k,t),  \nonumber \\ 
\langle\eta(k,t)\eta(-k,t')\rangle &=&2(\gamma_b T_b + \nu k^2
T_g)\delta(t-t'), 
\label{eq:langeq}
\end{eqnarray}
with $\langle\eta(k,t)\rangle=0$.  The amplitudes of the external and
internal noises in the above equation are reasonably assumed to be
proportional to $T_b$ and $T_g$, respectively.  The p.d.f. of
$v_\perp(k)$ measured in experiments are Gaussian, in agreement
with Eq.~(\ref{eq:langeq}), whereas those of real space
velocities $v_i$ show deviations from Gaussianity due to the relevance
of collisions in Eq.~(\ref{langgas}).
From Eq.~(\ref{eq:langeq}) we immediately obtain
\begin{equation}
nS_\perp(k)=N^{-1}\langle |v_\perp(k)|^2\rangle = \frac{\gamma_b T_b + \nu k^2 T_g}{\gamma_b + \nu k^2 }.
\label{eq:Sklowk} 
\end{equation}  
At equilibrium, namely when collisions are elastic and 
$T_g=T_b$, equipartition between modes is perfectly
fulfilled and the structure factor becomes flat,
i.e. $S_\perp(k)=T_b$.  Differently, in the granular case, where $T_g
\ne T_b$, equipartition breaks down and from Eq.~(\ref{eq:Sklowk})
we have that $S_\perp \to T_b$ for small $k$
and $S_\perp \to T_g$ for large $k$.
The Fourier inversion of Eq.~(\ref{eq:Sklowk}) in two dimensions yields:
\begin{equation}
nG_\perp({\mathbf r}) 
= T_g \delta^{(2)}({\mathbf r}) + (T_b- T_g)\frac{K_0(r/\xi)}{\xi^2},
\label{eq:Sx} 
\end{equation} 
where $K_0(x)$ is the 2nd kind modified Bessel function that, for
large distances, decays exponentially
\begin{equation}
K_0(r/\xi)\approx \sqrt{\frac{\pi}{2}}\frac{e^{-r/\xi}}{(r/\xi)^{1/2}},
\label{eq:Sx2} 
\end{equation} 
and we have introduced $\xi=\sqrt{\nu/\gamma_b}$. Such quantity turns
out to be a \emph{non-equilibrium} correlation length, measurable in
the system \emph{only} when non-conservative interactions are present.
Indeed, at equilibrium, the second term in the r.h.s. of
Eq.~(\ref{eq:Sx}) vanishes and no coherence can be identified in the
velocity field fluctuations.

The saturation of $n S_\perp(k)$ to a finite value at small $k$ provided
from this theory is in agreement with the behaviour of experimental
data observed in Fig.~\ref{Skexp}, suggesting 
that the model with finite friction is appropriate to describe the system.

\begin{figure}[!htb]
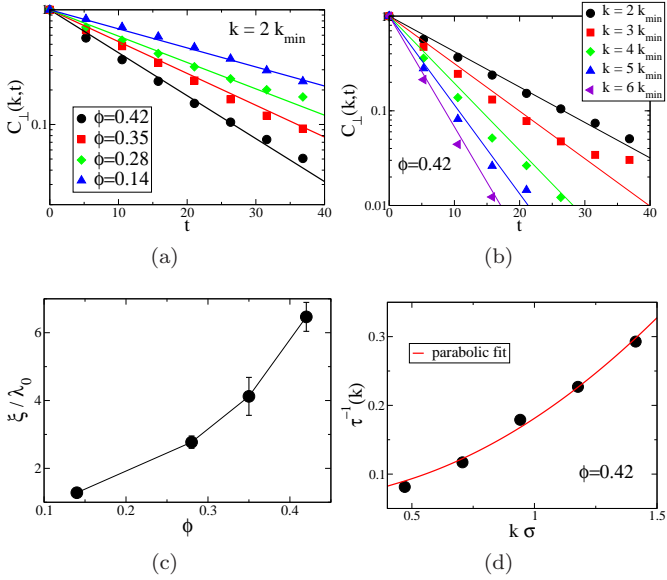

\begin{center}
\subfigure[]{\label{fig2:a}\includegraphics[width=0.49\columnwidth,clip=true]{figure3a.eps}}
\subfigure[]{\label{fig2:b}\includegraphics[width=0.49\columnwidth,clip=true]{figure3b.eps}} \\
\subfigure[]{\label{fig2:c}\includegraphics[width=0.49\columnwidth,clip=true]{figure3c.eps}}
\subfigure[]{\label{fig2:d}\includegraphics[width=0.49\columnwidth,clip=true]{figure3d.eps}}
\end{center}
\vspace{-0.7cm}
\caption{(color online). Data from experiments. (a) Temporal decay of
  $C_\perp(k,t)$ at fixed $k=2k_{min}$ and (b) for different $k$ at
  fixed $\phi=0.42$. The unit of times is $2\pi/\omega$. (c)
  Correlation length $\xi$, rescaled by $\lambda_0$, plotted as
  function of $\phi$. (d) Parabolic fit of $\tau^{-1}(k)=\gamma_b+\nu
  k^2$.}
 \label{fig:expdecay}
\end{figure}

The correlation length $\xi$ is defined through $\gamma_b$ and $\nu$,
which can be obtained by studying the dynamical correlation
$C_\perp(k,t)=\langle v_\perp(k,t)v_\perp(-k,0)\rangle/\langle
|v_\perp(k,0)|^2\rangle$.  Indeed, from Eq.~(\ref{eq:langeq}), we
immediately obtain $C_\perp(k,t)=\exp\{-(\gamma_b+\nu k^2)t\}$.  Such
behaviour is in good agreement with experimental data, as reported in
Fig.~\ref{fig:expdecay}, where $C_\perp(k,t)$ is plotted for several
values of $\phi$ (Fig~\ref{fig2:a}) and different wave-vectors $k$
(Fig.~\ref{fig2:b}), together with the best fit curves. At each
packing fraction, the length scale $\xi=\sqrt{\nu/\gamma_b}$ can be
obtained by comparing the decay of shear modes at different $k$ (see
Fig.~\ref{fig2:d}). {A parabolic fit
  $\tau^{-1}(k)=\gamma_b+\nu k^2$ shows that our data are broadly
  consistent with the theory, but it is also important to notice that
  only few data points are available.}  In Fig.~\ref{fig2:c} we see
how the correlation length $\xi$, rescaled with the
{prediction - in the diluted approximation - for the} mean
free path $\lambda_0= 1/(2\sqrt{\pi}g_2n\sigma)$, where
$g_2=(1-7\phi/16)/(1-\phi)^2$, significantly grows with the packing
fraction and signals an increasing degree of order in the granular
fluid. {Notice that the main contribution to the growth
  presented in Fig.~\ref{fig2:c} is due to the decrease of
  $\lambda_0$, while $\xi$ grows only weakly, as already visible in
  Fig.~\ref{Skexp}, where $S_\perp(k)$ has a ``sigmoidal'' shape with
  an inflection point almost independent of $\phi$. Such an
  observation is consistent with the experimental measurements
  discussed in~\cite{PEU02}. The growth of $\xi/\lambda_0$ denotes an
  increasing number of correlated particles.}

The finite extent of velocity correlations in a granular fluid was
already pointed out in~\cite{PEU02}. Here we discover that it is
strictly related to a saturation of $S_\perp(k)$ as $k \to 0$,
corresponding to a well defined energy of the ``external thermostat''.
Within the homogeneous cooling regime a length-scale $\xi$ was discussed
in~\cite{NEBO97}, but in that context it separated the stable,
$k>\xi^{-1}$, from the unstable $k<\xi^{-1}$, modulations of the
velocity field.

The above discussion shows that a fluctuation on the mode $k$ decays
with characteristic time $\tau(k)=1/(\gamma_b+\nu k^2)$. This
timescale has to be much larger than the microscopic time scale
given by the mean collision time $\tau_c\sim \lambda_0/\sqrt{T_g}$,
for the FHD with white noise to be effective. Comparing $\tau(k)$ and
$\tau_c$, we see that in our experiments such constraint is fulfilled
for a narrow interval of small $k$.  This range of validity is
constrained by parameters such as the kinematic viscosity $\nu$ and
the drag $\gamma_b$ of the thermal bath, which are not under control
in experiments. 

The predictions of the linearized FHD described above are tested by
simulating an interacting system of inelastic hard disk modelized by
Eq.~(\ref{langgas}), where $\gamma_b$ enters as a fixed parameter. In
particular, $\gamma_b^{-1}\sim \tau_b$ is the timescale of the thermal
bath, which can be tuned in order to have smaller values of
$T_g$. Indeed, by lowering $\gamma_b$, also $\nu$ is lowered, because,
in first approximation, $\nu\sim T_g^{1/2}$~\cite{GLB06}. Therefore,
by tuning $\gamma_b$, we can raise the shear mode decay timescale
$\tau(k)=1/(\gamma_b+\nu k^2)$, extending the range of validity of
FHD, i.e. the interval of $k$ values where Eqs.~(\ref{eq:langeq})
and~(\ref{eq:Sklowk}) are expected to hold.  We have performed 2D
{noisy} event-driven molecular dynamics simulations {of inelastic
  hard disks}, at different packing fractions $\phi\in [0.1,0.5]$,
with $\alpha\in [0.6,1]$.  {Details and results of the
  simulations are reported in a separate
  publication~\cite{GSVP11}}. We consider $N=10000$ disks in a box of
area $L^2$ with periodic boundary conditions and the packing fraction
is varied by changing the side length of the box.  The particles have
diameter $\sigma=0.01$ and unit mass and the thermal bath of
Eq.~(\ref{langgas}) has parameters $T_b=1$ and $\gamma_b=1$, while
$\tau_c$ is always in the range $[0.005,0.04]$.
\begin{figure}[!htb]
\includegraphics[width=1.0\columnwidth,clip=true]{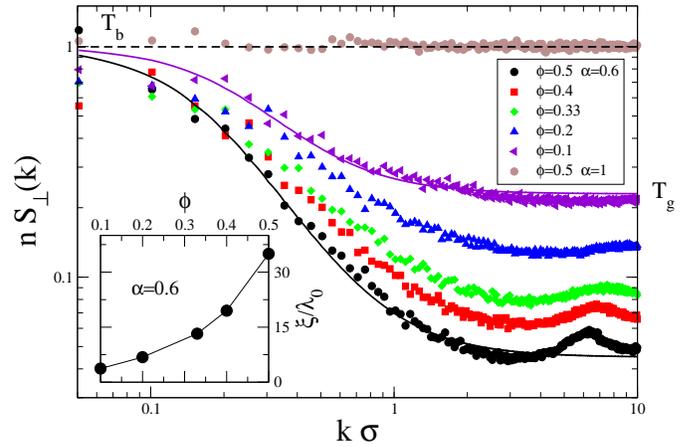}
\caption{(color online). $S_\perp(k)$ measured in event driven molecular 
dynamics simulations, averaged over 400 frames. 
The solid lines are the best fit curves via
Eq.~(\ref{eq:Sklowk}), for $\phi=0.42$ and $\phi=0.14$. Inset:
  $\xi/\lambda_0$ as a function of $\phi$.}
\label{Sknum}
\end{figure}
In Fig.~\ref{Sknum} we report our results for $S_\perp(k)$. A striking
analogy with the experimental results of Fig.~\ref{Skexp} can be
appreciated. Again, two energy scales are clearly observed: at small
$k$ all curves saturate to the value $T_b=1$, in agreement with
Eq.~(\ref{eq:Sklowk}).  Moreover, as expected, for $\alpha=1$ energy
equipartition holds and $S_\perp(k)=T_b$, whereas for $\alpha\ne 1$
the structure factor at large $k$ oscillates around the granular
temperature $T_g$.  Notice that here, by using Eq.~(\ref{eq:Sklowk}),
we obtained a good fit of data, in a wide range of $k$, thus
validating the FHD description (solid lines in Fig.~\ref{Sknum}).  The
kinematic viscosity $\nu$ obtained from these fits, yields the
correlation length $\xi$.  A remarkable increase
of $\xi/\lambda_0$ as function of $\phi$ is found (see the
inset of Fig.~(\ref{Sknum})).  Notice also that the gap between the
granular temperatures at different packing fractions and the bath
temperature is strongly increased, compared to the one observed in
experiments. Such a large separation between the two energy scales is
consistent with an extended realm of validity of the FHD.

In conclusion we have measured the transverse velocity structure
factor, and its time decay, for a $2D$ driven granular fluid. We have
observed a striking agreement with simulations of a microscopic model with
stochastic driving and viscous friction, and with a fluctuating hydrodynamics 
theory with two noise sources at different temperatures $T_b\ge T_g$. 
A central issue of our study has been to show the increasing extent of correlations
in the velocity field, with growing density. Such a phenomenon emerges in the system 
only when time-reversal symmetry is
broken, i.e. in the presence of an energy flux from the external driving to the
system, due to the inequality $T_g<T_b$.

\begin{acknowledgments} 
  We thank U.M.B.~Marconi and A.~Vulpiani for useful discussions and
  MD.~Deen for help in building the experimental setup.  The work is
  supported by the ``Granular-Chaos'' project, funded by the Italian
  MIUR under the FIRB-IDEAS grant number RBID08Z9JE.
\end{acknowledgments}


\bibliographystyle{eplbib}
\bibliography{fluct.bib}

\begin{thebibliography}{10}
\expandafter\ifx\csname url\endcsname\relax\def\url#1{\texttt{#1}}\fi

\bibitem{F75}
\Name{Foster D.} \Book{Hydrodynamic {F}luctuations, {B}roken {S}ymmetry, and
  {Correlation} {F}unctions} (Perseus {B}ooks) 1975.

\bibitem{JKPJ07}
\Name{J\"ulicher F., Kruse K., Prost J. \and Joanny J.-F.} \REVIEW{Phys. Rep.
  }{449}{2007}{3}.

\bibitem{NRM07}
\Name{Narayan V., Ramaswamy S. \and Menon N.} \REVIEW{Science
  }{317}{2007}{105}.

\bibitem{Cetal10}
\Name{Cavagna A. \etal} \REVIEW{Proc. Natl. Acad. Sci. USA }{107}{2010}{11865}.

\bibitem{JN92}
\Name{Jaeger H.~M. \and Nagel S.~R.} \REVIEW{Science }{255}{1992}{1523}.

\bibitem{DMB05}
\Name{Dauchot O., Marty G. \and Biroli G.} \REVIEW{Phys. Rev. Lett.
  }{95}{2005}{265701}.

\bibitem{K99}
\Name{Kadanoff L.~P.} \REVIEW{Rev. Mod. Phys. }{71}{1999}{435}.

\bibitem{G99}
\Name{Goldhirsch I.} \REVIEW{Chaos }{9}{1999}{659}.

\bibitem{GZ93}
\Name{Goldhirsch I. \and Zanetti G.} \REVIEW{Phys. Rev. Lett.
  }{70}{1993}{1619}.

\bibitem{NEBO97}
\Name{van Noije T. P.~C., Ernst M.~H., Brito R. \and Orza J. A.~G.}
  \REVIEW{Phys. Rev. Lett. }{79}{1997}{411}.

\bibitem{GLS90}
\Name{Grinstein G., Lee D.~H. \and Sachdev S.} \REVIEW{Phys. Rev. Lett.
  }{64}{1990}{1927}.

\bibitem{KSZ10}
\Name{Till W., Sperl K.~M. \and Zippelius A.} \REVIEW{Phys. Rev. Lett.
  }{104}{2010}{225701}.

\bibitem{VAZ11}
\Name{Vollmayr-Lee K., Aspelmeier T. \and Zippelius A.} \REVIEW{Phys. Rev. E
  }{83}{2011}{001301}.

\bibitem{BGL89}
\Name{Barrat J.~L., Gotze W. \and Latz A.} \REVIEW{J. Phys.: Condens. Matter
  }{1}{1989}{7163}.

\bibitem{SVGP10}
\Name{Sarracino A., Villamaina D., Gradenigo G. \and Puglisi A.}
  \REVIEW{Europhys. Lett. }{92}{2010}{34001}.

\bibitem{PLMPV98}
\Name{Puglisi A., Loreto V., Marconi U. M.~B., Petri A. \and Vulpiani A.}
  \REVIEW{Phys. Rev. Lett. }{81}{1998}{3848}.

\bibitem{PEU02}
\Name{Prevost A., Egolf D.~A. \and Urbach J.~S.} \REVIEW{Phys. Rev. Lett.
  }{89}{2002}{084301}.

\bibitem{RIS07a}
\Name{Reis P.~M., Ingale R.~A. \and Shattuck M.~D.} \REVIEW{Phys. Rev. Lett.
  }{98}{2007}{188301}.

\bibitem{RIS07b}
\Name{Reis P.~M., Ingale R.~A. \and Shattuck M.~D.} \REVIEW{Phys. Rev. E
  }{75}{2007}{051311}.

\bibitem{BC86}
\Name{Baus M. \and Colot J.~L.} \REVIEW{J. Phys. C: Solid State Phys
  }{19}{1986}{L643}.

\bibitem{NETP99}
\Name{van Noije T. P.~C., Ernst M.~H., Trizac E. \and Pagonabarraga I.}
  \REVIEW{Phys. Rev. E }{59}{1999}{4326}.

\bibitem{BMG09}
\Name{Brey J.~J., Maynar P. \and de~Soria M. I.~G.} \REVIEW{Phys. Rev. E
  }{79}{2009}{051305}.

\bibitem{GLB06}
\Name{Garc\'ia-Rojo R., Luding S. \and Brey J.~J.} \REVIEW{Phys. Rev. E
  }{74}{2006}{061305}.

\bibitem{GSVP11}
\Name{Gradenigo G., Sarracino A., Villamaina D. \and Puglisi A.} \REVIEW{J.
  Stat. Mech. }{}{2011}{P08017}.

\end{thebibliography}

\end{document}